\documentclass[journal, 12pt, onecolumn, draftclsnofoot]{IEEEtran}
\IEEEoverridecommandlockouts
\usepackage{cite}
\usepackage{caption}
\usepackage{amsmath,amssymb,amsfonts}
\usepackage{algorithmic}
\usepackage[ruled,vlined]{algorithm2e}
\usepackage{graphicx}
\usepackage{textcomp}
\usepackage{xcolor}
\def\BibTeX{{\rm B\kern-.05em{\sc i\kern-.025em b}\kern-.08em
    T\kern-.1667em\lower.7ex\hbox{E}\kern-.125emX}}

\begin{document}

\title{Design of AoI-Aware 5G Uplink Scheduler Using Reinforcement Learning\\
}

\author{\IEEEauthorblockN{Chien-Cheng Wu, Petar Popovski, Zheng-Hua Tan, Čedomir Stefanović}\\
\IEEEauthorblockA{Department of Electronic Systems, Aalborg University, Denmark}\\
\{ccw, petarp, zt, cs\}@es.aau.dk}

\maketitle

\begin{abstract}
	Age of Information (AoI) reflects the time that is elapsed from the generation of a packet by a 5G user equipment (UE) to the reception of the packet by a controller.
	A design of an AoI-aware radio resource scheduler for UEs via reinforcement learning is proposed in this paper.
	In this paper, we consider a remote control environment in which a number of UEs are transmitting time-sensitive measurements to a remote controller.
	We consider the AoI minimization problem and formulate the problem as a trade-off between minimizing the sum of the expected AoI of all UEs and maximizing the throughput of the network.
	Inspired by the success of machine learning in solving large networking problems at low complexity, we develop a reinforcement learning-based method to solve the formulated problem.
	We used the state-of-the-art proximal policy optimization algorithm to solve this problem.
	Our simulation results show that the proposed algorithm outperforms the considered baselines in terms of minimizing the expected AoI while maintaining the network throughput.
\end{abstract}


\section{Introduction}
Many emerging services and systems such as Autonomous Driving, Industrial Automation, and Tactile Internet require real-time monitoring and low-latency information delivery.
The growth of time-sensitive information led to a new data freshness measure named Age of Information (AoI). 
AoI evaluates the time elapsed from the generation of the last update at a source that was received at the destination~\cite{b1}.
Consider a cyber-physical system such as an automated factory where many robots 
are transmitting time-sensitive information to a remote
observer through a wireless network.
Each robot continuously samples the environment at the physical factory and transmits the sample data to the monitoring observer. 
Due to inevitable bandwidth limitations, the network may not be able to transmit all the data to the observer.
Consequently, each robot may be allocated a part of the transmission bandwidth that does not completely satisfy its requests.
In such cases, the latest generated data can not be transmitted immediately and will be accumulated at the robot's queue, 
resulting in the AoI increase.
A way to mitigate such scenarios is to 
optimize the operation of the network scheduler.

In the literature, many works are investigating the time-average AoI~\cite{b1} and aims to achieve the minimum time-average AoI.
If the network traffic from each node is known, the optimal scheduling policy for minimizing time-average AoI can be derived with low computation complexity~\cite{b15}.
However, the computation complexity of finding an optimal scheduling policy without any prior knowledge of the network traffic is high.

In this paper, leveraging the success of machine learning in solving many of the online large-scale networking problems \cite{b6}, we propose a method based on Reinforcement Learning (RL) to solve an optimization problem that combines network utility (a measure of matching between the required and allocated resources per user) and minimization of AoI.
In the training phase, the RL agent starts to capture network state in terms of the number of user equipments (UEs) and traffic volume of each UE. 
Then, at the given state, the RL agent is trained to choose a scheduling action that maximizes the total reward.
The RL agent continuously interacts with the environment and tries to find the best policy based on the reward fed back from the environment.
In the testing phase, obtain a policy capable of optimizing the transmission schedule.

As already hinted, the reward of the minimization problem considered in the paper combines two objective functions.
The first is the expected AoI of each node (i.e., user) and the second is the utility of each node.
These two objectives can be juxtaposed, as maximizing utility may adversely affect AoI and vice versa.
For example, consider a network with two nodes A and B, sharing the same wireless channel that can accommodate $L > 1$ packets per time slot. 
Node A and node B have a fixed size queue of length $L$.
Assume that node A generates $L$ packets per timeslot, while node B only generates $1$ packet per timeslot.
The scheduling policy selects which node is allowed to transmit packets at a given timeslot.
Policy $p1$ alternates between node A and node B.
Policy $p2$ selects node B every $L$ slots and selects node A otherwise.
Policy $p2$ has larger throughput than Policy $p1$ (i.e., every slot is fully used). 
But Policy $p1$ has lower AoI than Policy $p2$, as node B waits less for the transmission opportunities.
In the paper, we consider a generalization of the problem, in which the nodes are generating traffic with an a priori unknown statistics, 
and design an agent to learn the scheduling strategy that maximizes the reward based on the RL paradigm.

The rest of this paper is organized as follows.
Section~\ref{sec:related_work} presents a brief overview of the related work.
The system model and the problem formulation are described in Section~\ref{sec:sys_model}.
The proposed learning algorithm is presented in Section~\ref{sec:model_algo}, whereas the numerical simulations are given in Section~\ref{sec:simulation}. 
Finally, the conclusions are drawn in Section~\ref{sec:conclusion}.

\section{Background and Related Work}
\label{sec:related_work}

Recently, a number of papers tackled the problem of minimizing the AoI of many sources that are competing for the available radio resources. \cite{b15} considers the problem of many
sensors connected wirelessly to a single monitoring node and formulate an optimization problem that minimizes the weighted expected AoI of the sensors at the monitoring node.
The authors of~\cite{b16} also consider the sum expected AoI minimization problem when constraints on the packet deadlines are imposed.
In \cite{b17}, the minimization of the sum expected AoI
is considered in cognitive shared access. 

The scheduling decisions with multiple receivers over a perfect channel are investigated in \cite{b18,b11}, where the goal is to learn data arrival statistics.
Q-learning is used for a generate-at-will model in \cite{b18}, while policy gradients and DQN methods are used for a queue-based multi-flow AoI-optimal scheduling problem in~\cite{b11}. 
In addition, AoI in multi-user networks has been studied in \cite{b18}–\cite{b24}. 
The authors in~\cite{b21} show that finding an optimal scheduling decision that minimizes AoI is an NP-hard problem.

Scheduling transmissions to multiple receivers is investigated in~\cite{b18}, focusing on a perfect transmission medium, and the optimal scheduling algorithm is shown to be of threshold type on the AoI. 
Average AoI has also been studied when status updates are transmitted over unreliable multiple-access channels~\cite{b22} or multicast networks~\cite{b23}.
In addition, Peak AoI has also been jointly considered with Average AoI in~\cite{b37} and a UAV’s trajectory and average peak AoI optimization problem has been studied at~\cite{b35}.
A source node sending time-sensitive information to several users through unreliable channels is considered in~\cite{b20}, where the problem is formulated as a multi-armed bandit (MAB), and a suboptimal Whittle Index (WI) policy is proposed.

Most literature working on AoI minimization problems assumes perfect statistical knowledge of the random processes governing the status update system.
However, in most practical systems (e.g., heterogeneous UEs with different mission-critical traffic co-located in the same network), the characteristics of the system, e.g., mobility pattern, traffic distribution, etc, are not known a priori and must be learned. 
A limited number of recent works consider the unknown or time-varying characteristics of status update systems, and apply a learning-theoretic approach~\cite{b25,b11}.
To the best of our knowledge, the scheduling for a tradeoff between average AoI minimization and throughput maximization via utility modeling is studied for the first time at a multi-user system.

\section{System Model}
\label{sec:sys_model}

\begin{figure}
    \centering
	\captionsetup{justification=centering}
    \textbf{Mission-Critical System}\par\medskip
    \includegraphics[scale=0.6]{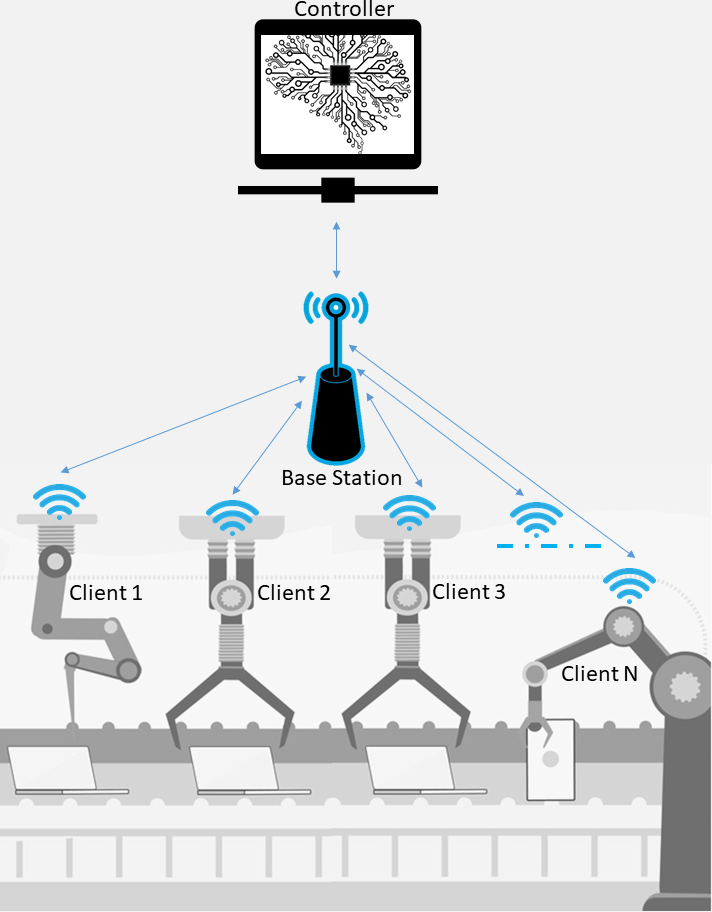}
    \caption{An autonomous control factory}
	\label{fig:architecture}
\end{figure}

In this paper, we consider a mission-critical system consisting of a 5G network with one base station (gNB), $N$ UEs and UE's controller as illustrated in Fig.~\ref{fig:architecture}.
The figure depicts a centralized autonomous-control factory, which is a mission-critical system.
The centralized controller monitors the state of each UE/robot through the 5G wireless network at the remote side.
The robots connect to the controller via the base station.
The connection between a robot and the controller can be modeled as a virtual link.
Since the robots briskly operate at the production line, to keep the system reacting in time, robots shall transmit fresh data such as sensor information or velocity to the controller and fetch control signals back to maintain the system operation.
All the UE-generated data can be carried in one or multiple packets and transmitted to the controller individually via the wireless link.
For the packet transmission scheduling over a link, a time-slotted system is considered, where scheduling decisions are made and transmitted to the UEs at the beginning of each time slot $t$.
Each time slot has a duration of $\tau$, e.g., 1 ms or even smaller by using dynamic transmission time intervals.
The total channel bandwidth is limited to $B$ units, where $0 < B < N$. 
Due to the network resource limitation, the network only allows a subset of UEs (denote as $m$, $ 0 < m \leq N$) at a timeslot $t$ to send packets to the central controller.
Denote by $S_{n}(t)$ be a random variable that indicates the selection of UE $n$ by the gNB at time slot $t$, i.e., $S_{n}(t) =1$ if the controller selects UE $n$  at time slot $t$, and $S_{n}(t) =0$  otherwise.
When $S_{n}(t) =1$, UE $n$ sends a subset of the packets, depending on the amount of allocated resource, to the controller. 
Otherwise, the UE cannot send any packets to the controller.
If the user (say user $n$) is selected for transmission at slot $t$, the allocation of channel bandwidth at time slot $t$ to UE $n$ is denoted by $b_n(t)$, where $b_n(t)$ is an integer number of units, $0< b_n(t) \leq B$ and $\sum_{n \in \mathcal{I}(t)} b_n (t) = B$, where $\mathcal{I}(t)$ is the set of selected users in slot $t$. 
The schedule at moment $t$ is denoted by $\mathcal{S}(t) = \{ S_1(t), S_2(t), \dots S_n(t), \dots S_N(t) \}$, which belongs to  the set of all possible schedules $\mathbb{S}$.

The number of packets generated at time slot $t$ by user $n$ is denoted by $X_{n}(t)$, following a memoryless arrival process that is independent and identically distributed over users.
The generated packets are stored at UE transmitters’ queues, following the first-come-first-serve (FCFS) policy.
The queue length is fixed and initialized to 0.

The AoI of each UE $n$ is defined as the time elapsed between the current time and the generation of the latest packet that departed from the transmitter; we assume that all the packets that are transmitted are also successfully received.
The AoI is computed by the formula:
\begin{equation} 
\label{age_def}
	A_{n}(t) := t-\underset{i}{\max}{\{t_{G}^{n}(i))|t_{G}^{n}(i)) \leq t\}}
\end{equation}
where $t_{G}^{n}(i))$ and $t_{D}^{n}(i)$ are the moments of generation and departure of the $i$th packet of UE $n$. 
Due to the slotted time assumption, AoI is changing in integer units (i.e., number of slots). 
If no packet from UE $n$ is received by the controller in time slot $t$, $A_{n}(t)$ is increased by 1.
Otherwise, if at time slot $t$, a packet from UE $n$ is successfully received by the controller, the AoI will be updated below:
\begin{equation}
\label{age_calc}
		{A}_n(t+1) = \begin{cases}
		{A}_n(t)+1 & \text{ if } {X}_n(t){b}_n(t){S}_n(t)=0
		\\ 
		t-t^n_G (i) & \text{ if } {X}_n(t){b}_n(t){S}_n(t)>1
		\end{cases}
\end{equation}
where $t^n_G (i)$ is the generation moment of the latest packet that was successfully received in slot $t$ (denoted by the dummy index $i$). 


Further, the node and network utility per slot are defined, respectively, as: 
\begin{align}
U_n(t) & = \frac{1}{1+e^{-(1.5 \times b_{n}(t) - X_{n}(t))}} S_{n}(t), \; 0 \leq n \leq N  \\
U(t) & = \sum_{n=1}^{N} U_n (t)
\end{align}
$U_n (t)$ is a sigmoid function, while 
the network utility $U(t)$ is the summation of all node utility values at time slot $t$.
The higher network utility means a higher network throughput~\cite{b37}.

Our goal is to find suitable scheduling policies for a set of UEs, which {attempt to maximize network utilization while minimizing the aggregate AoI. 
Specifically, we formulate our goal via the function given below.
In each time slot $t$, 
select a schedule $ \mathcal{S}(t)$  
s.t.                
\begin{align}
	\label{objective_function}
\mathcal{S}(t) = \mathop{\text{argmax}}_{\mathcal{S}(t) \in \mathbb{S}}\sum_{n=1}^{N} \left[ U_n (t)-\beta A_{n}(t) \right]
\end{align}
where $\beta \in \mathbb{R}$ is a scalar control parameter.
If $\beta$ is larger, the scheduling will be more sensitive than a small $\beta$.
Problem \eqref{objective_function} is a non-linear integer programming (NLIP) that is generally complicated to solve~\cite{b30}.
Actually, the optimization variables in \eqref{objective_function} include sequential decisions. 
It was recently shown that Deep RL (DRL) achieves high performance on the long-term sequential decision-making problems without human knowledge~\cite{b32,b33,b34}.
Moreover, DRL-based solutions can provide the decisions in an automatic and zero-touch manner. 
In addition, benefiting from deep neural networks, DRL is capable of handling high-dimensional observation-action spaces. 
These motivate us to propose policy-based model-free DRL solutions, discussed in the next section.

\section{Proposed Machine Learning Model and Algorithm}
\label{sec:model_algo}

We assume an RL agent interacting with a network environment to learn a scheduling policy without prior information.\footnote{If the base station has prior information such as the incoming traffic from the UEs or the AoI evolution, the base station scheduler can find the optimal scheduling policy by conventional algorithms.}
To deal with the unknown information in the network environment, an RL agent gradually learns the scheduling policy from the network environment observations.
We define the scheduling policy as $\pi(o_{t}, \theta_{t})$ that returns a schedule $\mathcal{S}(t)$ as an action $a_{t}$ to satisfy $\sum_{n=1}^{N} b_n(t) \leq B$.
The action $a_{t}$ represents which UEs are selected to transmit their packets. 
For example, if only $S_{1}(t) = S_{2}(t) = 1$ in $\mathcal{S}(t)$, it means that only UE1 and UE2 are selected to transmit data with their allocated bandwidth $b_1(t)$ and $b_2(t)$.
The other UEs are not allowed to transmit data and have to wait for the next action.
In the time-slotted system, every interaction between the agent and the network environment happens at the beginning of a time slot.
In each interaction, the agent samples the environment to get an observation $o_{t}$ and performs an action $a_{t}$ based on the scheduling policy.
After performing the action, the agent receives a reward $r_{t}$. Then the agent waits for the time slot $t+1$ to interact with the network environment.
The learning process repeats the interactions continuously to approximate the optimal scheduling policy which obtains the maximal reward. 
Hence, the objective of learning is to maximize the expected cumulative reward.

The optimization goal is defined as 
\begin{equation}
	J := \max\limits_{\pi_{k}} \mathbb{E} \left[ \sum_{n=1}^{N} \left( U_n (t) -\beta A_{n}(t) \right) \right]
\end{equation}
\begin{center}
s.t. $\sum_{n=1}^{N}b_n(t) \leq B$, $\pi_{k} = {\mathcal{S}(t)}$ \\
\end{center}
Our RL agent exploits the Proximal Policy Optimization (PPO) Algorithm~\cite{b32,b33}.
This PPO algorithm has become one of the most widely used algorithms in RL due to its better sample efficiency than other tabular-based RL algorithms.

\begin{figure}
    \centering
    \textbf{Reinforcement Learning Configuration}\par\medskip
    \includegraphics[scale=0.4]{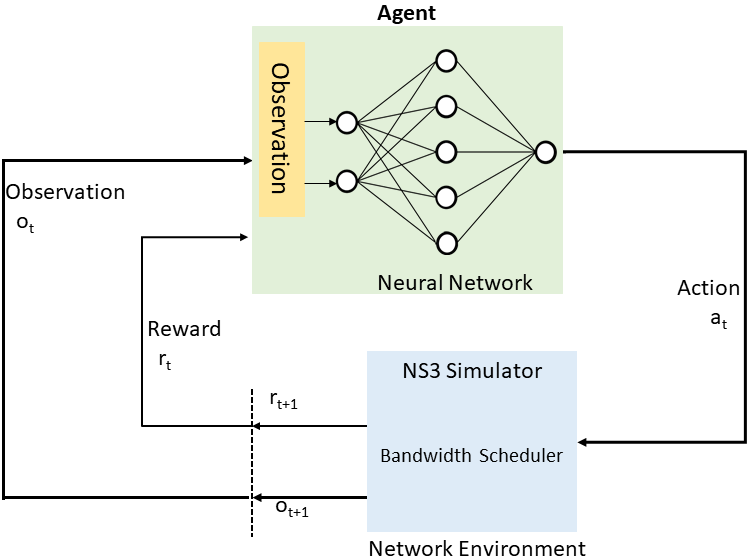}
    \caption{A Flowchart of the Reinforcement Learning Method}
	\label{fig:rl}
\end{figure}

As shown in Fig. \ref{fig:rl}, the learning process is a finite loop of $K$ iterations.
In the beginning, the agent will fetch an initial environment observation, $o_{t=0} = \{o_{1}(t=0)), o_{2}(t=0)), \dots o_{n}(t=0))\} $, from the network environment.
The agent observes a set of metrics from $o_{n}(t)$ including the buffer status, AoI value of every UE, and the throughput achieved in the last $k$ iterations. 
Then the agent feeds these values to the neural network, which will output the next action.
The next action is defined by which UEs are to be chosen for the next iteration $k+1$, as well as how much bandwidth they will be allocated, at time slot $t+1$. 
The scheduling policy is transformed from the action obtained from the trained neural network.
If a UE is selected to transmit packets then the corresponding bandwidth will be reserved for the UE. 
After the new scheduling is deployed to all the UEs, a reward is observed and fed back to the agent.
The agent uses the reward information to train and improve its neural network model.


Our implementation of the PPO algorithm in the scheduling problem is detailed in Algorithm \ref{algo:ppo}.
Starting from the initial parameters, the PPO algorithm optimizes its policy, $\pi$, until converges or reaches $K$ iterations.
At each iteration $k$, the PPO agent collects observation of a time slot. Next, it selects an action with the current policy.
After the agent takes the selected action, the agent obtains a reward based on the reward function. The reward function is defined as $R(o_{t}, a_{t})$.
In addition, we formulate a Q-value function, a value function and an advantage function which use to compute the intermediate values in each iteration:
\begin{equation}
	Q_{\pi}(o_{t}, a_{t}) = \mathbb{E}_{o_{t+1}, a_{t+1}}[ \gamma^{t} R(o_{t+1}) | o_{t} = o_{0}]
\end{equation}
\begin{equation}
	V_{\pi}(o_{t}) = \mathbb{E}_{o_{t+1}, a_{t}}[ \gamma^{t} R(o_{t+1}) | o_{t} = o_{0}]
\end{equation}
where $\gamma$ is a discount factor, $\gamma \in [0,1]$, and
\begin{equation}
	A_{\pi}(o_{t}, a_{t}) = Q_{\pi}(o_{t}, a_{t}) - V_{\pi}(o_{t})
\end{equation}

Then we construct the surrogate loss on these observations and optimize policy with SGD for $e$ epochs and minibatch size $\mathcal{B}$.

\begin{algorithm}
	\SetAlgoLined
	\DontPrintSemicolon
	\KwIn{An initial policy with parameters $\theta_{0}$ and initial observation $o_{0}$}
	\For{$k= 1, 2, 3, \cdots$ until $k= K$ or convergence}{
		Update age and bandwidth request based on observation $o_{k}$.\\
		Take scheduling action using policy $\pi=\pi(\theta_{k})$.\\
		Compute advantage estimation based on the value function.\\
		Optimize surrogate function $\nabla J$ with respect to $\theta_k$ using $e$ epochs and minibatch size $\mathcal{B}$.\\
		$\theta_k \leftarrow \theta_{k+1}$\\
	}
	\caption{Proximal Policy Optimization Algorithm}
	\label{algo:ppo}
\end{algorithm}

\section{Simulation Results and Discussions}
\label{sec:simulation}

In this section, we provide illustrate the performance of the scheme proposed in Section~\ref{sec:model_algo}.
To evaluate the scheme 
in a realistic cellular network, the simulation is performed in the network simulator (NS3)~\cite{b9}.
In addition, we select the round-robin algorithm as baseline 1 and a proportional-fair algorithm as baseline 2~\cite{b2}.
The simulation parameters in NS3 are listed in Table~\ref{tab:paratable}.

\begin{table}[ht]
    \centering
    \caption{Simulation Parameters}
    \label{tab:paratable}
    \begin{tabular}{ |p{4cm}||p{2cm}|}
        \hline
        \hline
        Parameter Name& Value\\
        \hline
		Number of UEs $N$ & $20$ \\
		Slot Duration $\tau$ & $1$ ms \\
		Packet Size (d) & $2048$ bytes \\
		Numerology & $0$ \\
		Duplexing & TDD \\
		Bandwidth & $20$ MHz \\
        Transmission Power & $20$ dBm  \\
		Propagation Model & TR 38.901  \\
        \hline
   \end{tabular}
\end{table}

\subsection{Simulation Setup}
\label{ssec:setup}

We revised the NS3 LTE module to implement a 5G environment.
The modulation and coding scheme and the resource block allocation are chosen based on the standard \cite{b3,b5,b8,b10,b13}, and \cite{b15}.
Users are distributed uniformly in the service area of 80*80 meters, while the base station is placed at a fixed location 
in the service area.
The simulation time was chosen to be 900 seconds which ensures that enough training samples were collected.
For each UE n, the packet arrival rate is randomly set with a normal distribution which has a mean frequency range between [60Hz, 1300Hz].
The variance of packet arrival rate is 9000 Hz.
The packet distribution will only be used in the Proportional-Fair algorithm as a known parameter.
Regarding the parameters of the actor, 
we set the batch size $\mathcal{B} = 80$. 
Then, we set the step size as $0.6$ and use a three-layer DNN with hyperbolic tangent (Tanh) activation function, Adam optimizer, and initial learning rate $0.0003$.
On the other hand, the neural network structure of critic 
starts with an action-appended input layer.
Then, it connects to a fully connected hidden layer and an output layer with N outputs. 
The critic also uses the Tanh activation function and Adam optimizer. Additionally, the initial critic learning rate is set to $0.001$.

\subsection{Simulation Results}
\label{ssec:results}

The proposed scheme is assessed in terms of the average AoI and average UEs throughput. 
Based on those two metrics, the following scheduling schemes are compared:
\begin{enumerate}
	\item Round-Robin (RR), where all RBs are evenly allocated to each UE;
	\item Proportional-Fair (PF), where all RBs that are allocated depend on the known arrival traffic distribution;
	\item Proximal Policy Optimization (PPO), where all RBs are allocated by the prediction from our RL agent.
\end{enumerate}

We now present our simulation results based on the two baseline algorithms and the proposed PPO algorithm~\ref{algo:ppo}. 
Fig.~\ref{fig:aoi} shows the average AoI and Fig.~\ref{fig:throughput} presents the network throughput as functions of the mean traffic generation frequency.
The RR algorithm has the worst AoI performance and the lowest throughput due to the full fairness for each UE.
The PF algorithm has the best AoI and throughput performance because the data generating distributions over the network nodes are known parameters.
It can be seen that the PPO algorithm outperforms the round-robin algorithm at the heavy traffic condition without any proprietary parameters from the UEs.
Finally, in the PPO algorithm, the scheduler outperforms the RR algorithm and achieves almost as good performance as the PF algorithm, despite the lack of knowledge of the data generating distributions.

\begin{figure}
    \centering
    \includegraphics[scale=1.2]{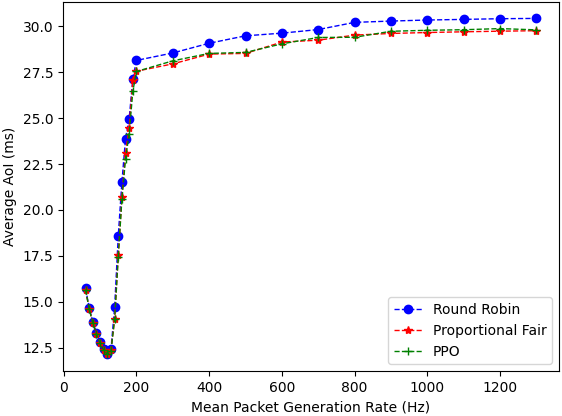}
    \caption{Average AoI in different Packet Generation Rates}
	\label{fig:aoi}
\end{figure}

\begin{figure}
    \centering
    \includegraphics[scale=1.2]{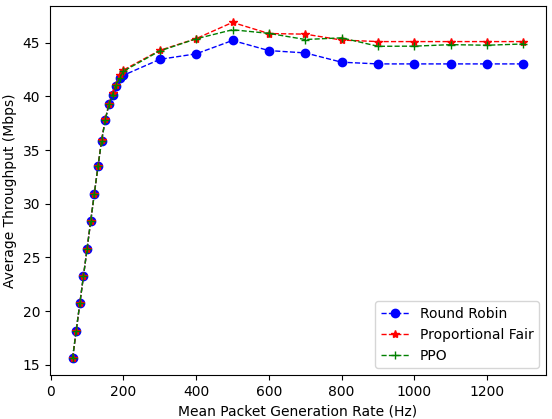}
    \caption{Average Throughput in different Packet Generation Rates}
	\label{fig:throughput}
\end{figure}

\subsection{Discussions}
\label{ssec:discussions}

If the network bandwidth is larger than the total traffic requirements, there is no network backlog.
Thus, any scheduler can achieve the same AoI performance.
Therefore, all the algorithms have the same AoI performance at low traffic conditions, as illustrated in Fig.~\ref{fig:aoi}.
However, when the total traffic requirements exceed the available bandwidth, the individual traffic from each UE starts backlogging, and AoI starts to increase.
This happens in our evaluation when the data generation rate surpasses $120$~Hz.
The RR algorithm has the worst AoI performance and the lowest throughput because the RR algorithm does not consider the individual traffic load.
The PF algorithm has better AoI and throughput performance due to the pre-configured traffic generation distribution. 
However, in the practical system, the traffic distribution may not be obtained.
It can be seen in Fig.~\ref{fig:aoi} and Fig.~\ref{fig:throughput} that the PPO algorithm achieves the high throughput while keeping the low AoI without the knowledge of the traffic distribution.
Since the PPO algorithm adapts to the traffic generation and AoI evolution,
the PPO agent flexibly allocates the network bandwidth. 

\section{Conclusions}
\label{sec:conclusion}

In this paper, we 
designed a model-free deep reinforcement learning (DRL) method for optimizing the uplink scheduler.
DRL agent guided the scheduler to deal with network utilization and UEs age minimization in a scenario with unknown traffic generation.
The problem formulation and optimization process of AoI provided a theoretical basis for future studies on next-generation network radio resource management.
Moreover, the proposed learning framework of centralized estimation and execution could be deployed in the real network environment.
For our future work, we will consider learning-based scheduling in heterogeneous network architectures.

\end{document}